\documentstyle[aps,prb,multicol,epsf]{revtex}

\newcommand{\beq}{\begin{equation}}
\newcommand{\eeq}{\end{equation}}
\newcommand{\bea}{\begin{eqnarray}}
\newcommand{\eea}{\end{eqnarray}}

\begin{document}

\title{Comment on the intermediate vortex liquid phase in extreme type-II
superconductors}

\author{ M.~V.~Feigel'man,}

\address{
L.~D.~Landau Institute  for  Theoretical  Physics,  117940  Moscow,
Russia
}

\maketitle

\begin{abstract}
It is shown that disentangled vortex liquid
can be a true thermodynamic phase of type-II superconductors
in spite of the absense of supercurrent in the
direction parallel to the magnetic field.

\end{abstract}

\begin{multicols}{2}

The problem of the nature of phase transition(s)
into a superconductive state in extreme type-II superconductors
placed in external magnetic field was recently addressed in a number
of numerical simulations\cite{tachiki,teitel,sudbo} and experiments
~\cite{samoilov,bariloche} on YBa$_2$Cu$_3$O$_7$ single crystals.
The main issue was formulated long ago~\cite{feigel1} as follows: does an
intermediate state of vortex matter ("disentangled vortex liquid", DVL)
 exist between well-understood
 Abrikosov lattice and normal metal states ?  Following early
 discussions in \cite{feigel1,feigel2}, the DVL  was understood
as the state with superconductive current response,
$J_z = - {\rm const}\cdot \delta A_z$, along the direction of
background magnetic field ${\bf B} \parallel \hat{z}$.
With such an understanding, the results of numerical and experimental
studies~\cite{tachiki,teitel,sudbo,samoilov,bariloche} are
apparently conflicting: measurements~\cite{samoilov} and
simulations~\cite{sudbo} indicate the presence of some intermediate
vortex liquid state, whereas other measurements~\cite{bariloche} and
simulations~\cite{tachiki,teitel,sudbo} demonstrate absense of
superconductive response in the vortex liquid.
In the present Comment I would like to point out a simple way to
reconcile the above observations and to provide some
understanding of the vortex state discussed in~\cite{samoilov,sudbo}.

Consider a 3D superconductor with current ${\bf J}$ and vector potential
${\bf A}$. The longitudinal
response of the superconductor in the vortex state is defined by,
\begin{equation}
J_z({\bf q},q_z)= -P_{zz}({\bf q}, q_z) A_z({\bf q}, q_z),
\end{equation}
where ${\bf q} \perp \hat{z}$.
There is a  ``duality relation'' between the
longitudinal polarization function of the  superconductor
$P_{zz}$
and the transverse polarization function
$\Pi_\perp(\omega =0, {\bf q})$
of the corresponding
2D Bose-liquid ground-state~\cite{feigel1,feigel2}
\beq
P_{zz}({\bf q}, q_z =0) = \frac{1}{\lambda^2}\frac{q^2}{q^2 +
g^2\Pi_\perp(\omega =0, {\bf q})},
\label{dual}
\eeq
(in the context of the frustrated XY model~\cite{teitel}, \,
$\Upsilon_z({\bf q})=P_{zz}({\bf q},q_z=0)$ is called
the helicity modulus).
For the superfluid groud-state of 2D Bose-lqiuid,
$\Pi_\perp(q\to 0) = \rho_s >0$,
 and the duality relation~(\ref{dual}) gives a normal longitudinal response
$P_{zz} \propto q^2$. For the  2D Bose-liquid ground-state with
finite susceptibility $\chi$, like in normal (non-superfluid) liquid,
one gets $\Pi_\perp({\bf q}\to 0) =\chi q^2$, and Eq.~(\ref{dual}) gives
$\Upsilon_z(q\to 0) > 0$, i.e. a non-zero helicity modulus.
Numerical data~\cite{tachiki,teitel,sudbo} show no evidence
of such a liquid state. The point to be made is that
these data {\it do not contradict} to the existence of non-superfluid
Bose-liquid ground-state {\it per se}.

Namely, suppose that a Bose-liquid
ground-state possesses a transverse
polarization function with an anomalous $q$-dependence,
$\Pi_\perp({\bf q}) \propto |q|^a$ with $ 0 < a < 2$.
The corresponding state of the superconductor will
have $\Upsilon_z (q\to 0) \propto |q|^{2-a} $, so the $q=0$ helicity
modulus is zero.  At the same time, such a ground-state is qualitatively
different from the superfluid Bose liquid (i.e. "entangled vortex liquid"
in superconductive language); its existence is intrinsically related
to the appearence of massless excitation modes with nonzero momentum  $q$,
and breakdown of Galilei invariance, as described in~\cite{susy}.
On the other hand, an existence of an intermediate state with
such a massless modes is the direct consequence of experimental
observations~\cite{samoilov}.
The only numerical data for the finite-$q$ helicity modulus
$\Upsilon(q)$ are provided in~\cite{teitel}, cf. their Fig.3b.
The data for $T=0.54$ seem to be
consistent with the above hypothesis, although their accuracy is not yet
sufficient to establish this.
An additional phase
transition above the melting line observed
(by counting of vortex lines trajectories )
 in Ref.~\onlinecite{sudbo} can be understood macroscopically
as a transition into a state {\it dual} to the
above non-superfluid Bose-liquid.

On a purely theoretical side, arguments in favor of existence
 of an intermediate nonsuperfluid Bose-liquid ground state
were advanced in Ref.~\onlinecite{susy} where a Bose liquid with a special
kind of instantaneous interaction was considered.
The peculiar feature of the interaction is that it produces no
long-wavelength shear modulus $C_{66}(q\to 0) = 0$ even in the
periodic crystal-like ground state;
as a result, the ground state was found to be very fragile.
 Although a physical
 realization of the specific model studied in Ref.~\onlinecite{susy} looks
doubtful, it has the same qualitative property,
$C_{66}(q\to 0) = 0$,  as the vortex-vortex interaction in the recently
studied borocarbide compounds of the RNi$_2$B$_2$C family at the critical
line $H^*(T)$ of the hexagonal-to-square vortex lattice transition.\cite{boro}
We speculate that there may be a fluctuation region around the mean-field
transition line $H^*(T)$
between the vortex lattices of different symmetries, and that
an anomalous vortex liquid state may occur
in borocarbides within this region.

To conclude, a simple way to reconcile apparently conflicting
published results~\cite{tachiki,teitel,sudbo,samoilov,bariloche}
 on existence of disentangled vortex liquid in HTSC is
proposed.
I am grateful to V.~B.~Geshkenbein and A.~E.~Koshelev for
 many useful discussions.
This research was supported by the
RFBR grant \#98-02-16252 and the DGA grant \#94-1189.

\end{multicols}

\end{document}